\def\simlt{\lower.5ex\hbox{$\; \buildrel < \over \sim \;$}}
\def\simgt{\lower.5ex\hbox{$\; \buildrel > \over \sim \;$}}
\RequirePackage{lineno} 
 \documentclass[preprint2]{emulateapj}
 \usepackage{graphicx}

\newcommand{\myemail}{mrline@ucsc.edu}
\slugcomment{Accepted to the Astrophysical Journal }
\shorttitle{Brown Dwarf Retrieval}
\shortauthors{Line et al.}

\begin{document}
\title{A Data-Driven Approach for Retrieving Temperatures and Abundances in Brown Dwarf Atmospheres}
\author{Michael R. Line}
\author{Jonathan J. Fortney}
\affil{Department of Astronomy and Astrophysics, University of California, Santa Cruz, CA 95064}
\author{Mark S. Marley}
\affil{NASA Ames Research Center, Mail Stop 245-3; Moffett Field, CA 94035}
\author{Satoko Sorahana}
\affil{Department of Physics, Nagoya University, Nagoya, Aichi 464-8602, Japan}
\email{mrline@ucsc.edu}
\altaffiltext{1}{Correspondence to be directed to \myemail}

\begin{abstract}
Brown dwarf spectra contain a wealth of information about their molecular abundances, temperature structure, and gravity.  We present a new data driven retrieval approach, previously used in planetary atmosphere studies, to extract the molecular abundances and temperature structure from brown dwarf spectra.  The approach makes few a priori physical assumptions about the state of the atmosphere.  The feasibility of the approach is first demonstrated on a synthetic brown dwarf spectrum.   Given typical spectral resolutions, wavelength coverage, and noise properties precisions of tens of percent can be obtained for the molecular abundances and 10s-100s K on the temperature profile.  The technique is then applied to the well studied brown dwarf, Gl 570D.  From this spectral retrieval the spectroscopic radius is constrained to be 0.75 - 0.83 $R_{\mathrm{J}}$, $\log(g)$ to be 5.13 - 5.46 and $T_{\mathrm{eff}}$ to be between 804 and 849 K.   Estimates for the range of abundances and allowed temperature profiles are also derived.  The results from our retrieval approach are in agreement with the self-consistent grid modeling results of Saumon et al (2006).  This new approach will allow us to address issues of compositional differences between brown dwarfs and possibly their formation environments, disequilibrium chemistry,  missing physics in current grid modeling approaches as well as a many other issues.

\end{abstract}

\section{Introduction}

Unlike most stars, with photospheres that predominantly emit over a limited range of altitudes, the molecular opacity-dominated atmospheres of brown dwarfs have large variations in opacity with wavelength, allowing flux from very different atmospheric depths to emerge over various spectral ranges. This property--which brown dwarfs share with planetary atmospheres--allows information to be extracted from a range of altitudes and conditions, if spectral measurements are available from a sufficiently broad swath of wavelengths. 

 Historically, by comparing models to observations, observed spectra have been interpreted to discern brown dwarf masses,  formation modes, evolution,  and the processes at work in their atmospheres (Burrows et al. 1993; Allard et al. 1996; Allard et al. 2001; Marley et al. 1996; Saumon et al. 2000;  Geballe et al. 2001;  Burrows et al. 2006; Hubeny \& Burrows 2007; Burgasser et al. 2007;  Cushing et al. 2008; Stephens et al. 2009; Rice et al. 2010; Yamamura et al. 2010; Patience et al. 2012).  These processes include vertical mixing, disequilibrium chemistry, global dynamics, and cloud formation and sedimentation.  Currently, spectra of brown dwarfs are generally interpreted through comparisons of observed spectra with pre-computed grids of model atmospheres.  Such models self-consistently solve for the temperature structure in radiative-convective equilibrium and molecular abundances in the atmosphere with few free parameters, typically $\log g$ and effective temperature, cloud parameterizations, and in some cases an eddy mixing coefficient to accommodate non-equilibrium chemistry due to vertical mixing.  The combination of $\log g$ and effective temperature that provide a best fit, and the subsequent atmospheric structure, are taken to be the solution for a particular brown dwarf's atmosphere.  Within the framework parameter grid search or Monte-Carlo methods are sometimes implemented in order to estimate the uncertainties in these parameters (e.g., Cushing et al. 2008; Rice et al. 2010).
 

While this grid-based comparison approach has offered considerable insight into interpretation of the spectra, these methods are constrained by various assumptions which do not easily allow for a full exploration of brown dwarf parameter space.  For instance, most self-consistent grid models assume thermochemical equilibrium, which we now know is generally not the case, especially with molecular species like NH$_3$/N$_2$ and CH$_4$/CO which can be driven strongly out of equilibrium due to vertical mixing (e.g., Saumon et al. 2006; Griffith \& Yelle 1999) .  Within the self consistent grid model frame work this is remedied by some via the inclusion 1D vertical mixing prescription parameterizied with an eddy diffusivity parameter, but such an approach is often reliant upon the choosing the correct rate limiting steps (e.g., see Visscher \& Moses 2011; Moses et al. 2011).   Furthermore, most of these models generally assume solar elemental abundances, such that modeling investigations of metal-poor and metal-rich atmospheres, or deviations from solar-like C/N/O ratios have generally been lacking (although see Allard 1997; Tsuji et al. 2011; Burrows et al. 2006; Saumon \& Marley 2008).  The underlying causes for deviations of best-fitting models from data have rarely been explored, even though such deviations record shortcomings in the underlying model assumptions or underlying physical data.  The vast majority of these models also assume 1D radiative-convective equilibrium in a static atmosphere, leaving one unable to explore departures from these temperature structures arising from dynamical transport of heat, disequilibrium chemistry, latent heat release due to condensation, or other phenomena. Furthermore, given the widespread observational evidence for variability in the emitted spectra of brown dwarfs, new modeling approaches that relax assumptions inherent in previous models are now needed.


The goal of this investigation is to introduce a new approach for the interpretation of brown dwarf spectra. Powerful techniques to directly invert measured spectra into constraints on molecular abundances and atmospheric temperature structure have been widely used within the Earth sciences (Rodgers 1976, Towmey 1996, Rodgers 2000, Crisp et al. 2004) and solar system planets (Conrath et al. 1998, Irwin et al. 2008, Nixon et al. 2007, Fletcher et al. 2007, Greathouse et al. 2011) and recently exoplanet atmosphere inference (Lee et al. 2012; 2013 Line et al. 2012;  Barstow et al. 2013; Line et al. 2013a).  These atmospheric retrieval approaches are primarily data driven and free from many of the aforementioned assumptions, and naturally allow for a wide exploration of brown dwarf atmosphere parameter space.  The atmospheric temperature structure and molecular abundances are ``retrieved" through the iterative calculation of many tens to thousands of model spectra.  Each spectrum is generated with a unique temperature structure and variations on the molecular abundances.  An assessment is made regarding the goodness of fit of each model.  Since an extremely large phase space of temperature structures and abundances are probed, the end product is a range of models (or an analytic estimate of that range) that fit the observed spectra, along with a statistical assessment of the goodness of fit from a wide range of models.

In the current paper we apply the above well established retrieval methodologies to brown dwarf spectra in order to illustrate the veracity of the approach. We first retrieve the thermal structure and atmospheric abundances of a model and then turn to the well-studied T dwarf Gl 570D, which has previous been the target of extensive observational (Burgasser 2000;2003;2006; Leggett et al. 2002; Cushing et al. 2006; Patten et al. 2006; Geballe et al. 2009) and modeling campaigns (Geballe et al. 2000; Saumon et al. 2006).  The paper is organized as follows:  In \S\ref{sec:Methods} we introduce the retrieval methodology.  A synthetic example is shown in \S\ref{sec:Synthetic} and our initial results for Gl570D are in \S\ref{sec:Gl570}.  Finally, in \S\ref{sec:Conclusions} future research directions are discussed.

\section{Methods}\label{sec:Methods}

Many statistical tools exist for the parameter estimation problem.  The most powerful of these tools falls under the umbrella of Bayesian statistics. Bayesian approaches make simultaneous use of the data and prior information.   One particularly prominent method, Optimal Estimation (e.g., Rodgers 2000) is briefly describe here (see Rodgers 2000 or Line et al. 2012 for further details).  This approach uses Bayes theorem to arrive at the following penalty/cost function:
\begin{eqnarray} \label{eq:cost_func}
\chi^2({\bf x})={({\bf y}-{\bf F(x)})^{T}{\bf S_{e}^{-1}}({\bf y}-{\bf F(x)})} \nonumber \\
+({\bf x}-{\bf x_{a}})^{T}{\bf S_{a}^{-1}}({\bf x}-{\bf x_{a}}) 
\end{eqnarray}
where $\bf y$ is the set of $n$ data points (the spectrum), $\bf x$ is the $m$-dimensional parameter state vector (in this work, a vector of temperatures, abundances, gravity and radius), $\bf F(x)$ is the forward model, and $\bf S_{e}$ is the $n \times n$ data error matrix who's diagonal elements are the square of the 1$\sigma$ data error bars and the off diagonal elements are set to zero in the absence of uncorrelated noise.   $\bf x_{a}$ is the {\em a priori} state vector and $\bf S_{a}$ is the $m \times m $ {\em a priori} covariance matrix who's diagonal elements are the square of the a-prior 1$\sigma$ uncertainties on the desired parameters.    These values are typically taken to be large in order to mitigate the influence of the prior on the retrieval.   The first term in equation \ref{eq:cost_func} is simply the standard ``chi-squared" and the second term represents the prior knowledge of the parameter distribution before we make the observations.  The prior represents our state of knowledge before we make the observations.  In this investigation, the prior is an encapsulation of what we think the atmospheric state (temperature profile, gas abundances, gravity etc.) should look like before making the observations.    The optimal solution is the one that maximizes the a-posteriori probability which is the equivalent of minimizing equation \ref{eq:cost_func}.  Standard numerical routines such as the Levenberg-Marquardt method can be used to minimize this often non-linear cost function.   The uncertainty on the retrieved state vector can be estimated with the following covariance matrix, which describes the hyper-dimensional Gaussian posterior distribution:
\begin{equation}\label{eq:shat}
{\bf \hat{S}=(K^{T}S_{e}^{-1}K+S_{a}^{-1})^{-1}}
\end{equation}
where {\bf K} is the jacobian matrix, or matrix of partial derivatives that describe the sensitivity of the forward model (change in flux at each wavelength) to each of the parameters in the state vector.  This error estimate makes use of the local hyper-dimensional gradient information ({\bf K}) and the data error bars to make a point estimate of the parameter uncertainties about the best fit.  This error estimation is valid in the presence of random noise and in a regime in which the posterior probability distributions can be assumed gaussian.  Based upon similar retrieval modeling of exoplanet data, Line et al. (2013b) showed that this error estimation approach is valid in moderate signal-to-noise and resolution regimes (R $\ge$ 100, SNRs $\ge$ a few ) typical of brown dwarf data, but breaks down for low signal-to-noise/resolution regimes.  

There are also other retrieval methodologies, including the bootstrap Monte Carlo approach and the Markov chain Monte Carlo (MCMC) approaches. The MCMC approach explores the posterior by running many thousands of forward models to evaluate the posterior probability.  This approach has been recently used in the exoplanet atmospheric retrieval problems (e.g. Madhusudhan et al. 2011; Benneke \& Seager 2012; Line et al. 2013). The bootstrap Monte Carlo approach (see \S \ref{sec:Gl570approach}) is a type of data resampling approach.  The approach resamples the data many thousands of times and re-applies the optimal estimation approach to each regenerated data sample in order to more accurately explore the uncertainty in each model parameter (also see Ford 2005).
 Line et al. 2013 showed that high quality data with a large number of spectral points and high resolution-as is brown dwarf data, that the full exploration of parameter space via MCMC methods may not be necessary and that the optimal estimation and bootstrap Monte Carlo approaches and their inherent assumptions can appropriately capture the uncertainties. With complicated and slow forward models such approaches may be advantages over the somewhat cumbersome and computationally demanding MCMC approaches.  We refer the reader to Line et al. (2013) for a detailed comparison of the various approaches and the appropriateness of the Gaussian posterior assumption within the context of extrasolar planet data.
 
 The most important component of a retrieval (regardless of the exact retrieval method) is the forward model.  We define the forward model as the component of the retrieval algorithm that creates model data given some input parameters (e.g., molecular abundances and temperature profiles) that can be directly compared to the measured data.  This is different than a 1D atmospheric structure model that self-consistently computes the temperature structure and molecular abundances under various assumptions.   The forward model has two tasks.  The first is to compute the model observation vector ({\bf F(x)}) of which can be compared to the data ({\bf y}) and the second is to compute the jacobian matrix ({\bf K}).  Our particular forward model is a modified version of the CHIMERA (Line et al. 2013) forward model which solves the thermal infrared radiative transfer equation to compute the disk-integrated thermal emission spectrum of an object scaled to ten parsecs as observed at Earth given a temperature profile, gas mixing ratio's, radius, distance, and gravitys'.  We also assume that the atmosphere is in hydrostatic equilibrium.  The high-resolution flux is convolved with the appropriate instrumental response functions for a given data set.  Clouds are not included in this preliminary investigation.  The jacobians are computed analytically as this improves accuracy and computational efficiency.  The analytic jacobians for the gas mixing ratios (H$_2$O, CH$_4$, CO, CO$_2$, and NH$_3$) and temperature at each level are described in Line et al. (2013) and Irwin et al. (2008).   The analytic jacobian that describes the intensity response to a change in gravity at a given wavelength is given by
 \begin{eqnarray}\label{eq:g_jac}
\frac{\partial I_{\lambda}}{\partial g}=-\sum_{z=0}^{N_{lev}}B_{\lambda}(T_{z})e^{-\sum_{j=z}^{N_{lev}}\Delta \tau_{j,\lambda}}\frac{\Delta \tau_{z,\lambda}}{g}\nonumber\\
 +\sum_{z=0}^{N_{lev}}(B_{\lambda}(T_{z})e^{-\sum_{j=z}^{N_{lev}}\Delta \tau_{j,\lambda}}\Delta \tau_{z,\lambda}\sum_{j=z}^{N_{lev}}\frac{\Delta \tau_{j,\lambda}}{g})
\end{eqnarray}
where $\Delta \tau_{j,\lambda}$ is the total optical depth of the $z^{th}$ slab, $g$ is gravity, and $B_{\lambda}(T_{z})$ is the blackbody emission at temperature $T_{z}$ in the $z^{th}$ slab.  The jacobian for a scaling factor representative of (R$_{p}$/R$_{I}$)$^{2}$ is also computed.

The 1-dimensional atmospheric structure is parameterized with variables that directly impact the emergent spectrum such as the gas abundances, temperature profile, gravity, and radius.  We aim to retrieve the mixing ratios of five gases: H$_2$O, CH$_4$, CO, CO$_2$, and NH$_3$.   H$_2$/He continuum absorption is also included where the H$_2$/He mole fraction (with He/H$_2$=0.193) is computed by subtracting the latter molecules from unity.    All mixing ratios are assumed to be uniform with altitude.  This assumption is valid in regimes in which vertical mixing homogenizes gas mixing ratios, though the impact of vertical mixing ratio structure should be explored in a future study.   The Freedman et al. (2008) cross section data base was used with the updates to the ammonia and H$_2$ collision-induced opacities described in Saumon et al. (2012).  For simplicity, alkali metals, metal oxides or hydrides are not included in this investigation.  We choose wavelengths over which these absorbers have minimal impact and hence, should not affect the retrieval.   The temperature profile is not parameterized, rather the temperature at each model level is retrieved.   However,  some smoothing is implemented through the a-prior covariance matrix in equation (\ref{eq:cost_func}) to prevent overfitting and unphysical oscillations in the profiles.  
  In the next section  this retrieval approach is applied to a synthetic data set of which the true atmospheric state (temperature structure and molecular mixing ratios) is known.

\begin{figure*}
\begin{center}
\includegraphics[width=0.75\textwidth, angle=0]{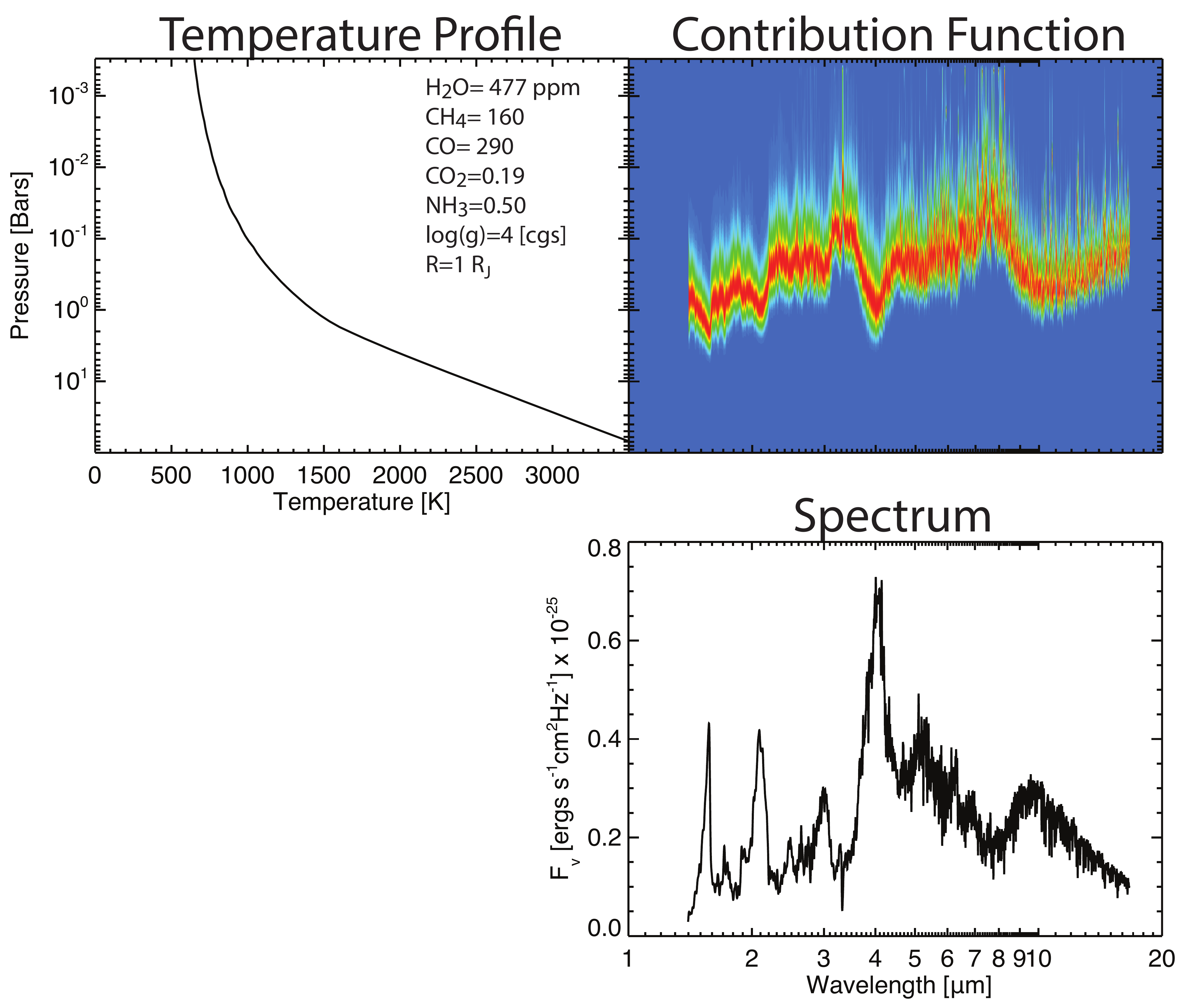}
\end{center}
     \caption{ \label{fig:Figure1} Synthetic brown dwarf atmosphere described in \S\ref{sec:Synthetic}.  The model atmosphere (temperature profile and gas mixing ratios) is shown in the top left panel.  The top right panel shows the normalized thermal emission contribution functions--from where in the atmosphere the emission originates.  Red indicates the peak of the contribution functions, where the optical depth is unity, and blue represents no contribution to the emergent spectrum.  Finally, the bottom right panel spectrum resulting from the given temperature profile, mixing ratios, and gravity shown in the top left panel  smoothed to a resolution of 0.01 $\mu$m. }
\end{figure*} 
 
\section{Synthetic Retrieval Analysis}\label{sec:Synthetic}
\begin{figure}
\includegraphics[width=0.45\textwidth, angle=0]{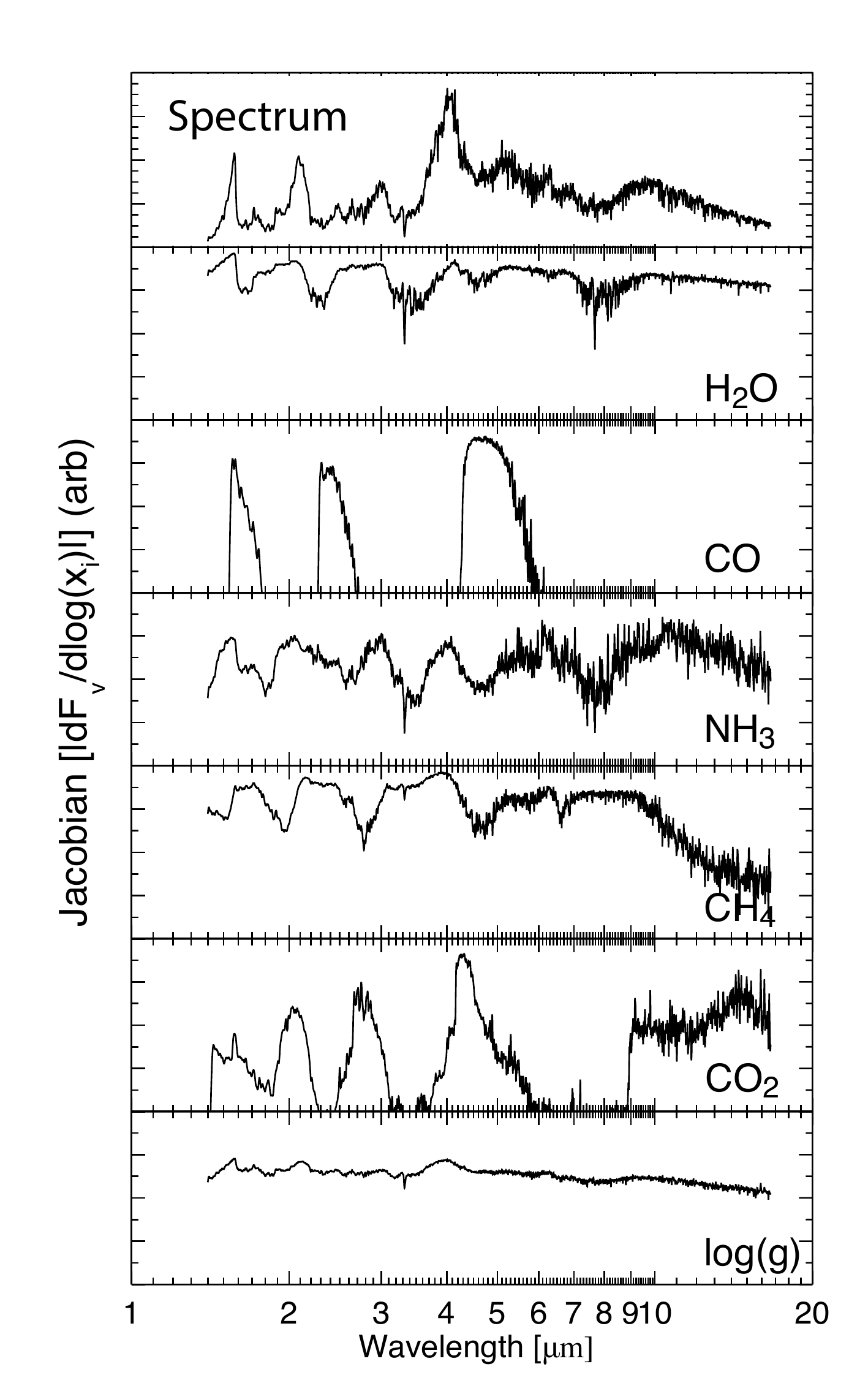}
     \caption{ \label{fig:Figure2} Gas and gravity jacobians.  The absolute value of the jacobians is shown in arbitrary units but all are on the same relative scale.  These show how the flux at every wavelength responds to an increase in value of the indicated parameter (x$_{i}$).  The jacobian is negative at all wavelengths (absolute value shown here) for all gases and positive at all wavelengths for gravity (see text).  The panel on the top is the same spectrum shown in Figure \ref{fig:Figure1} for comparison.  The greatest sensitivities occur in the strong bands of the molecules. }
\end{figure} 

In order to validate our approach, we first create a synthetic brown dwarf spectrum with artificial noise.  The synthetic brown dwarf is one Jupiter radius at 10 parsecs with a cloud free, $\log(g)$ of 4.0, $T_{\mathrm{eff}}$=1300 K temperature profile from a Saumon \& Marley (2008) grid model.  For simplicity uniform-with-altitude mixing ratios for H$_2$O, CH$_4$, CO, CO$_2$, and NH$_3$ are assumed.  Though the chosen values are somewhat arbitrary, they are similar to what one might expect assuming thermochemical equilibrium at solar composition at these temperatures and representative pressures (though these abundances may not necessarily be entirely consistent with the temperature profile).  We generate a high resolution spectrum from 1.5 -15 $\mu$m at  1 cm$^{-1}$ resolution ($\lambda/\Delta\lambda$=5000 at 2$\mu$m, 2000 at 5$\mu$m).  Again,  wavelengths shorter than 1.5 $\mu$m are not considered as this is where alkali metals, hydrides, and metal oxides present significant absorption.  

 Figure \ref{fig:Figure1} summarizes the synthetic brown dwarf atmosphere.  The contribution functions (pressure derivative of the transmittance times the local planck function) show that most of the emission originates between $\sim$2 and 10$^{-2}$ bars.  The H and K bands  and the 4 $\mu$m window see the deepest layers.  Regions of strong methane absorption, at 3.3 $\mu$m and 7.6 $\mu$m tend to probe the lowest pressures.
 
 As mentioned in \S\ref{sec:Methods}, the jacobian describes how the emergent spectrum at all wavelengths responds to a perturbation in one of the parameters.  This is the core information the retrieval uses to determine the optimal solution.  Figure \ref{fig:Figure2} summarizes the gas and gravity jacobians evaluated at the true atmospheric state used to compute the synthetic data.  These plots can be interpreted as the flux response spectrum to a change in order-of-magnitude of the parameter.   Generally speaking, increasing the gas abundance in an atmosphere with a temperature profile that monotonically decreases with altitude, will result in a negative flux response while increasing the objects gravity will produce a positive response.  This is because as the abundance of a gas increases, the peak of the contribution function moves towards lower pressures where the temperature is cooler, and hence a lower flux.  Increasing gravity, on a fixed temperature profile, pushes the peak of the contribution function to a deeper pressure (using the relation that $\tau\propto P/g$) where the atmosphere is hotter, hence more flux.
 
The model spectrum is now simulated as it would be observed through several instruments.  Here,  only photon limited random noise is assumed. Systematic biases due to photometric calibration errors or other sources have not been included.  The object was ``observed" under typical instrumental characteristics (e.g., Cushing et al. 2006; Geballe et al. 2009) with the NASA Infrared Telescope Facility (IRTF) SpeX instrument (Rayner et al. 2003) at a spectral resolution of 0.0025 $\mu$m ($\lambda/\Delta\lambda$=800 at 2$\mu$m) and signal-to-noise (SN) of 65 , an M-band spectrum at a resolution 0.01 $\mu$m ($\lambda/\Delta\lambda$=450 at 4.5$\mu$m) and a SN of 10, and finally the Spitzer Infrared Spectrometer (IRS, Houck et al. 2004) at a spectral resolution of 0.03 $\mu$m for $5.40 \le \lambda \le7.53 \mu$m ($\lambda/\Delta\lambda$=200 at 6$\mu$m) and 0.06 $\mu$m for $7.55 \le \lambda \le 14.63 \mu$m ($\lambda/\Delta\lambda$=167 at 10$\mu$m) and an SN of 10.   

\begin{figure*}
\begin{center}
\includegraphics[width=0.95\textwidth, angle=0]{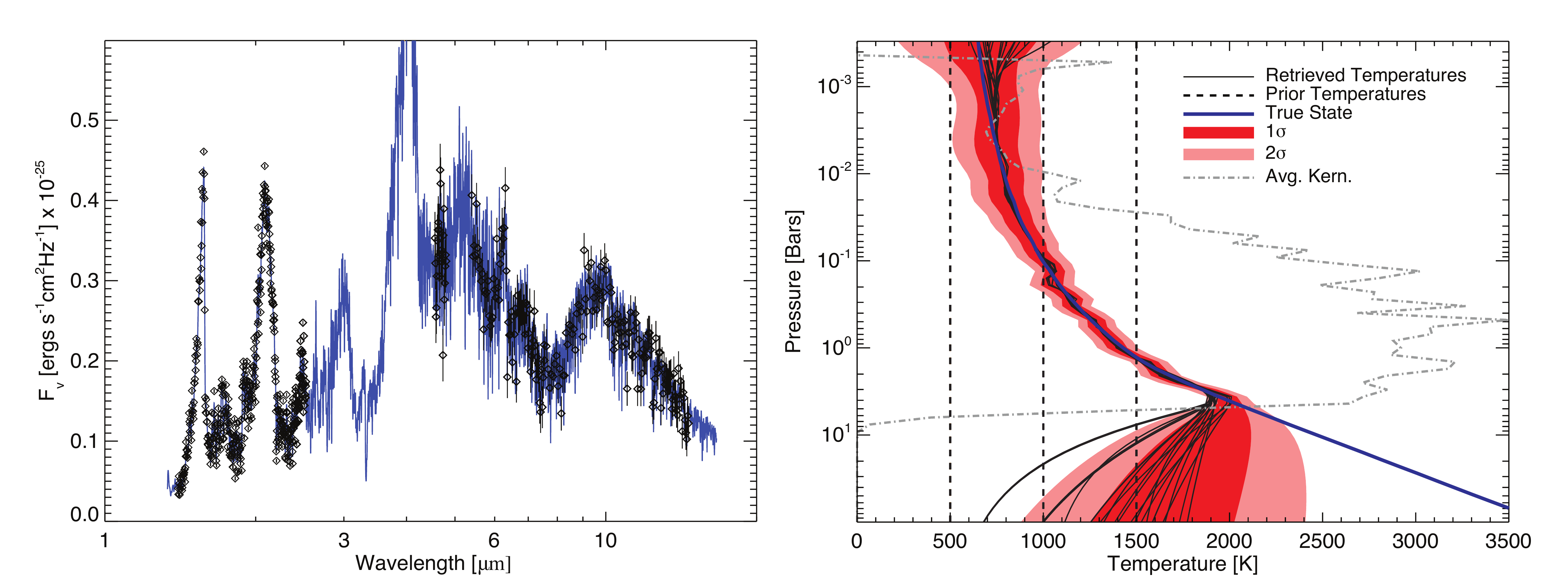}
\end{center}
     \caption{ \label{fig:Figure3} Synthetic spectrum with noise added (left) and resulting retrieved temperature profiles (right).  The simulated spectrum is shown with the black diamonds and the error bars.  A representative best fit spectrum is shown in blue.  In the right panel we summarize the temperature profile retrieval results.  The thick dark blue profile is the true profile.  The black vertical dashed lines are the temperature prior means that resulted in statically good fits.  The black temperature profiles are best fit profiles resulting from different prior assumptions.  The red envelope represents the 1 and 2$\sigma$ temperature profile uncertainty estimated from equation \ref{eq:shat} from a single best fit.  The profiles resulting from different prior assumptions all converge within the red uncertainties and to the true state (dark blue) suggesting a proper error estimation and an insensitivity to reasonable prior assumptions.  Furthermore, the profiles diverge from the true state below $\sim$4 bars.  This is because of a lack of spectral information from these deeper levels.   The gray curve summarizes the normalized averaging kernel which shows where the temperature profile can be well constrained.  Note that the temperature profiles diverge from the true state as averaging kernel goes towards zero.   }
\end{figure*}

We are now in a position to retrieve the temperature profile, gas mixing ratios (H$_2$O, CH$_4$, CO, CO$_2$, and NH$_3$), gravity, and radius from the simulated synthetic spectrum shown in first panel in Figure \ref{fig:Figure3}.  The parameters to be retrieved are the log of the uniform-with-altitude mixing ratios of the aforementioned 5 gases, $\log(g)$, radius, and 70 levels of the temperature profile for a total of 77 free parameters in the state vector.  The mixing ratios are, again, independent of the temperature profile and are not self-consistently computed using the assumptions of thermochemical equilibrium or quench chemistry. Realistically, when retrieving the radius we are retrieving a scale factor multiplying the flux.  If the distance is known, the radius can be obtained from this scale factor.  In our synthetic example, the distance is set to 10 pc.  

The retrieval process also requires that a prior be considered as defined by a prior mean, $\bf x_{a}$, and  prior covariance matrix,  $\bf S_{a}$.  Given there may be very little a prior knowledge of a newly discovered object it is important to explore the impact of a wide range of priors.  The prior also serves as the initial guess in the retrieval, though in general it need not necessarily be.  The selection of a wide range of priors also improves the odds of finding the global minimum as opposed to local minima in which non-linear minimization techniques tend to get trapped.  The largest impact on the retrieval will result from the temperature profile prior assumptions.  Because of this, many temperature profile priors are assumed, where the prior means are isothermal atmospheres at 500, 1000, 1500, 2000, 2500, and 3000 K.  Isothermal atmospheres are the simplest and perhaps most incorrect prior assumption that can be made.  A wide range of prior widths are chosen as well, 5, 10, 100, and 500K.  These widths represent the 1$\sigma$ range over which the temperature profile is expected to deviate from the isothermal prior means.   More physically motivated priors could have been chosen, however we prefer to demonstrate that the approach works with the simplest of temperature profile assumptions.  There is certainly not enough information content in one spectrum to independently retrieve the temperature at each atmospheric level.  In order to prevent overfitting of the data (which can result in unphysical oscillations in the temperature profile where there is little information) some smoothing of the temperature profile must be applied, known as Tikonov regularization, as implemented via a smoothing parameter in off diagonal elements of the  $\bf S_{a}$ matrix.  These off diagonal elements are given by equation 2.83 in Rodgers (2000) or equation 12 in Line et al. (2013).  This smoothing effectively reduces the total number of independent levels that must be retrieved.  The smoothing parameter describes over how many scale heights the temperatures are correlated.  A range of smoothing values, 1.5, 3, 5, and 7, are explored.  
 
We also explore the impact that the gravity prior has on the results.   $\log(g)$ prior mean values of 3, 4, and 5 are chosen.  These values encompass the likely range of $\log(g)$ in brown dwarf atmospheres.  The gravity prior widths are 1.5 orders of magnitude.   The gas or radius priors are not varied.  The gas prior means are chosen are somewhat close to what one would expect thermochemically.  The widths are broad however, thus mitigating their impact.  The radius prior width is also allowed to span 30$\%$ on either side of the mean.   The combinations of all of the aforementioned priors gives a total of 288 different priors.
 
\begin{figure*}
\begin{center}
\includegraphics[width=0.75\textwidth, angle=0]{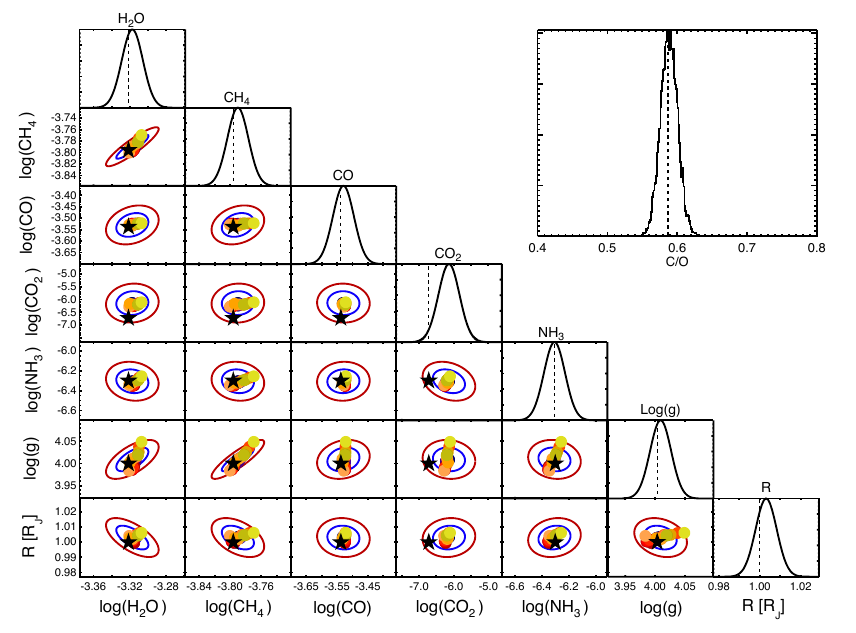}
\end{center}
     \caption{ \label{fig:Figure4} Gas (in log mixing ratio), gravity (in log gravity in cgs), and radius (Jupiter radii) retrieval uncertainties.  This ``stair-step" plot represents the Gaussian posterior distributions for each parameter derived from the covariance matrix given in equation \ref{eq:shat} for these parameters.  The gas abundances are given by the log of their volume mixing ratios.  The blue and red curves are the 1 and 2$\sigma$ error ellipses.  The gaussians on the top of each column represent the marginalized uncertainty for that parameter.  They are all Gaussian to do the assumptions inherent in the optimal estimation retrieval approach.  The vertical dashed lines and the star are the true state.  The colored circles are the retrieved values resulting from the different priors that fit the data well.  They all fall within the error ellipses suggesting that the Gaussian assumption made in the optimal estimation approach is reasonable for this scenario.  The angle of the ellipses shows the degree of correlation of one parameter versus the other.   The inset shows the derived C to O ratio distribution resulting from the retrieval uncertainties.    }
\end{figure*}

 Due to the nature of minimizing algorithms, many of the retrievals resulting from the 288 different priors can get trapped in local minimum resulting in fairly poor fits.  Thus the reasoning for exploring many different priors/initial guesses.  Only the retrievals that have the lowest values of equation \ref{eq:cost_func}, namely, ones that result in a cost-function per number of data points of less than two, are retained.  This means that only retrievals that result in fits to within the 2$\sigma$  data error bars (assuming the first term in equation \ref{eq:cost_func} dominates due to the large prior widths) are retained.  In several cases, since the model perfectly describes the synthetic data set (e.g., no missing model physics or systematic biases),  near perfect fits are attained.  Of the 288 priors/initializations, only 30 (10$\%$) meet this criterion.  It is these results that are highlighted in Figures \ref{fig:Figure4} and \ref{fig:Figure5}.  
 
 The right panel in Figure \ref{fig:Figure3} summarizes the retrieved temperature profiles.  We do not use the ensemble of retrieved results to derive the parameter uncertainties, rather we choose one single best fit and estimate the retrieval uncertainty with equation \ref{eq:shat}.  The red envelope shows the resultant one and two sigma uncertainty.  This uncertainty envelope encompasses the true state (blue) and all of the retrieved temperature profiles resulting from the remaining 29 priors.  This demonstrates that the error estimation via equation \ref{eq:shat} is believable.  Note that the temperature profiles all converge to nearly the same solution regardless of the prior assumptions.  This means that the data intrinsically has enough information content to constrain the temperature-pressure profile independent of the prior assumptions.
 
 Another diagnostic used to assess the believability of the retrieval is the averaging kernel.  The averaging kernel is an $m$-parameter by $m$-parameter matrix that describes how the retrieved state of one parameter depends on the true state of another parameter.  Perhaps more usefully, the diagonal elements of this matrix describe how much fractional information in constraining a particular parameter came from the prior versus the data (Rodgers 2000, Line et al. 2013), or in other words, how much you should believe the result.     When the diagonal element of the averaging kernel matrix for particular parameter is unity, that means all of the information used to constrain that parameter came from the data.  If it is zero, that means that the data did not inform the retrieval and the parameter value and its uncertainty stems solely from the prior.  The trace of the averaging kernel matrix gives the total number of degrees of freedom, or the number of independent pieces of information that can be retrieved from the data.  This value will be equal to the total number of free parameters in the model if the data is perfect and the model is well matched to that data.
 
 The gray curve shown with the temperature profiles in Figure \ref{fig:Figure3}  shows the normalized diagonal elements of the averaging kernel matrix for each of the temperature levels.  It can almost be thought of as an ``averaged" thermal emission contribution function.  In this particular case, the temperature profile is most well constrained between $\sim4$ and 10$^{-2}$ bars, roughly the same span of the contribution functions in Figure \ref{fig:Figure1}.  Regions above and below this are more strongly influenced by the prior suggesting these results should be interpreted with some caution.  This is clearly seen at pressure levels deeper than 4 bars.  Notice how the temperature uncertainty envelope grows substantially and how all of the retrieved profiles below this level relax towards their prior mean states, far from the true state.  Essentially, there is absolutely no temperature information in this spectrum from pressure levels deeper than 4 bars, simply because there are few spectral features that have thermal emission contribution functions in these regions of the atmosphere.   Retrieval precisions of better than $\sim$50 K are achieved over the atmospheric regions of which were are sensitive.  By computing the trace of the averaging kernel over the temperature profile parameters we find that there are 14 independent pieces of temperature profile information that can be retrieved, with most of the information (13 pieces of information) concentrated between $\sim4$ and 10$^{-2}$ bars.  
 
 Finally, we show the retrieved gas abundances and uncertainties in the form of a covariance plot in Figure \ref{fig:Figure4}, as determined by equation \ref{eq:shat}.  The blue and red ellipses represent the 1 and 2$\sigma$ uncertainties, respectively, for a single best fit.  Like the temperature profile, the retrieval encompasses the true state (star) within the uncertainties.  Furthermore, the retrieved parameter values (colored circles) resulting from the other 29 priors fall within the error ellipses.  These combined suggest that the retrieval result  and uncertainty estimation is both precise and accurate and do not depend on the priors.   In fact the retrieved precisions are better than 10$\%$ for H$_2$O, CH$_4$, CO, $\log(g)$, and radius.  CO$_2$ and NH$_3$ are retrieved to better than a factor of 2.  This is un-precedented precision, on par with what can be done on solar system objects.  For perspective, with the best observed exoplanet (HD189733b, see Madhusudhan \& Seager 2009; Lee et al. 2012; Line et al. 2014) precisions to no better than an order-of-magnitude (1000$\%$) can be obtained.   
 
 With the derived abundances it is instructive to compute the elemental carbon-to-oxygen ratio. It is hypothesized that objects that form in proto-planetary discs beyond the water or CO ice lines may have a carbon enriched gaseous envelope (e.g., {\"O}berg et al. 2011, Helling et al. 2014).  Since many brown dwarfs presumably form in molecular clouds they will have envelopes that do not show such a carbon enhancement.  The C to O ratio is derived by summing the total number of carbon atoms in the carbon carrying species by the total number of oxygen atoms in the oxygen bearing species following formula:
\begin{equation}
	 \mathrm{C/O=\frac{\Sigma C}{\Sigma O}=\frac{CH_4+CO+CO_2}{H_2O+CO+2CO_2}}.
\end{equation}
 The C to O ratio derived from the synthetic retrieval results is shown as the inset in Figure \ref{fig:Figure4}. The high precision on the gas abundances results in a precise C to O determination.  With such a high precision an enhanced C to O ratio of unity can be ruled out by $\sim$30$\sigma$. 
 
 
In this section we have demonstrated the potential for brown dwarf data to constrain temperatures and molecular abundances to unprecedented precision.  Such precision will be useful in constraining bulk atmospheric properties such as the metallicity and C to O ratios.  Furthermore, the high temperature precision may allow us to determine variability driven by thermal processes.  However, this precision is owed to the fact that the synthetic data only contained random noise and that the model was well matched to the data.  Real data is plagued by systematic uncertainties due to photometric calibration issues or other sources, and generally the model may not be well matched to the data.  For instance, absorption cross-sections could be in-accurate, or the assumption of uniform-with-altitude mixing ratios may not be true.  We will explore some of these issues when applying our retrieval to the well studied brown dwarf, Gl 570D.

\section{Retrieval Analysis of Gl 570D}\label{sec:Gl570}

The late T-dwarf, Gl 570D (Burgasser et al. 2000)is a benchmark system that has a well characterized spectral energy distribution and well known system properties (metallicity, distance, and age).  Geballe et al. (2001) used the optical and near IR spectrum along with the system constraints to accurately determine the effective temperature and gravity.  Saumon et al. (2006) and Geballe et al. (2009) put further constraints on the atmospheric composition/chemistry by making use of both near and mid infrared data covering $\sim$0.6 to 15 $\mu$m. They found that strong vertical transport of ammonia and CO are required to explain anomalous absorption in the M-band and IRS data.

We revisit Gl 570D with our newly developed retrieval methodology in order to determine the atmospheric molecular abundances, gravity, and spectroscopic radius.   This late T-dwarf is deliberately chosen due to its presumed cloud free nature which significantly reduces there required forward model complexity.   Evolutionary models are not needed to to constrain any parameters.   Datasets from three different instruments are combined covering 1.1 - 15 $\mu$m.  These are IRTF SpeX (Burgasser 2006) from 1.1 -  2.4 $\mu$m (R$\sim$300 at 1.5 $\mu$m), AKARI (Sorahana \& Yamamura 2012) from 2.5 - 5 $\mu$m (R$\sim$415 at 4 $\mu$m), and Spitzer IRS (Cushing et al. 2006) from 5.5 - 15 $\mu$m (R$\sim$165 at 10 $\mu$m) (see Figure \ref{fig:Figure5}).  This is the first analysis combining all three of these datasets to derive atmospheric properties of Gl 570D.  The data are photometrically calibrated and rescaled to a distance of 10 parsecs (actual distance of 5.91 parsecs, Perryman et al. (1997)).  This broad wavelength coverage encompasses many of the molecular absorption features (H$_2$O, CO, CO$_2$, CH$_4$, and NH$_3$)  expected to be present at Gl 570D's cool effective temperature (T$\sim$800 K).   The SpeX data short ward of 1.1 $\mu$m is clipped in order to obviate the need to include alkali and metal hydride opacity sources in the forward model.  By clipping the data short ward of 1.1 $\mu$m, we lose a small amount of information on the temperature structure, and cannot derive alkali abundances.  Neglecting these species should have little impact on other retrieved gas abundances as gases have many spectral features long wards of 1.1 $\mu$m that do not overlap with the alkali metals.  We will investigate the  alkali metals and their impact on the retrieved temperature structure in a separate investigation.   The region between 1.63 and 1.97 $\mu$m is also discarded.  Historically poor fits of models to data in this spectral range have been attributed to uncertainties in the methane line list (e.g., Cushing et al. 2008) and comparisons of various available line lists continue to show a wide disparity in this region. Rather than introduce an additional source of uncertainty, here we simply neglect this spectral region in the SpeX data. In the future we will further explore model sensitivity to methane opacity in this region.

\begin{table}
\centering
\caption{\label{tab:Table1} The 68\% confidence for each of the Gl 570D parameters from our retrievals compared to those of S06.  The mixing ratios quoted from S06 come from the 800 K level (see Figure \ref{fig:Figure7}).  }
\begin{tabular}{ccc}
\hline
\hline
\cline{1-2}
Parameter &	This Work & S06    \\
\hline
Radius (R$_{J}$)& 0.75 - 0.83	& 0.85 - 0.90	\\
log(g) (cm s$^{-2}$)&5.13 - 5.46	& 5.09 - 5.23	\\
M (M$_{J}$)   &   35 - 74		&  38 - 47	\\
T$_{eff}$ (K)&804 - 849	&800 - 821	\\
H$_2$O	& 8.25$\times10^{-04}$  - 1.60$\times10^{-03}$	&  7.6$\times10^{-04}$		\\
CH$_4$	& 3.32$\times10^{-04}$ -  5.43$\times10^{-04}$	& 5.0$\times10^{-04}$		\\
CO	& 1.35$\times10^{-05}$ -  3.65$\times10^{-05}$	& 1.7$\times10^{-06}$		\\
CO$_2$	& 9.52$\times10^{-08}$ -  2.47$\times10^{-07}$	& 9.4$\times10^{-12}$	\\
NH$_3$	& 1.28$\times10^{-06}$  - 7.34$\times10^{-06}$	& 1.3$\times10^{-05}$		\\
C/O	&  0.136 - 0.235	& Solar (0.55)		\\
\hline
\end{tabular}
\end{table}

\subsection{Approach}\label{sec:Gl570approach}
We undergo a slightly different approach than in \S\ref{sec:Synthetic} in order to accommodate for systematic errors due to the uncertainties in the photometric calibrations.  Such normalization errors are not readily accommodated within the optimal estimation framework used above.  First the optimal estimation retrieval approach is applied, ignoring the systematic uncertainties for now,  just as in \S\ref{sec:Synthetic}.  Then the prior that produces a fit that results in the smallest value of equation \ref{eq:cost_func} is identified.  Finally a data resampling method (Bootstrap Monte Carlo, Press et al. 1995; Ford 2005) is applied in order to better characterize the uncertainties in the retrieval parameters while accommodating for the photometric calibration error.  Approximately 10000 resampled data sets are generated by drawing a new flux value at each wavelength bin from a gaussian distribution with a width given by the gaussian noise data error bar.   Each of the three data sets are then multiplied by a scale factor (on the order of unity) that is also drawn from a gaussian distribution with a width given by the photometric calibration uncertainty in each data set as reported by the authors.  This effectively simulates any photometric ``jitter" due to the calibration uncertainties.  Photometric uncertainties are considered to be the only systematic uncertainties present in the data.    The reported photometric uncertainties for the SpeX, AKARI, and IRS data are 5$\%$, 10$\%$, and 6.5$\%$, respectively.  These calibration uncertainties can be larger than the non-systematic gaussian noise on each data point.  The optimal estimation approach, with the optimal prior, is then applied to each of these data realizations.  This data resampling approach is similar to the approach taken in Rice et al. (2010) and Geballe et al. (2009).   A hard upper limit on $\log(g)$ is enforced preventing values above 5.5 as values larger than this are not permitted by evolution models (e.g., Saumon \& Marley 2008), though this upper limit may be relaxed.

\subsection{Results \& Comparison to Saumon et al. (2006)}
\begin{figure*}
\begin{center}
\includegraphics[width=0.95\textwidth, angle=0]{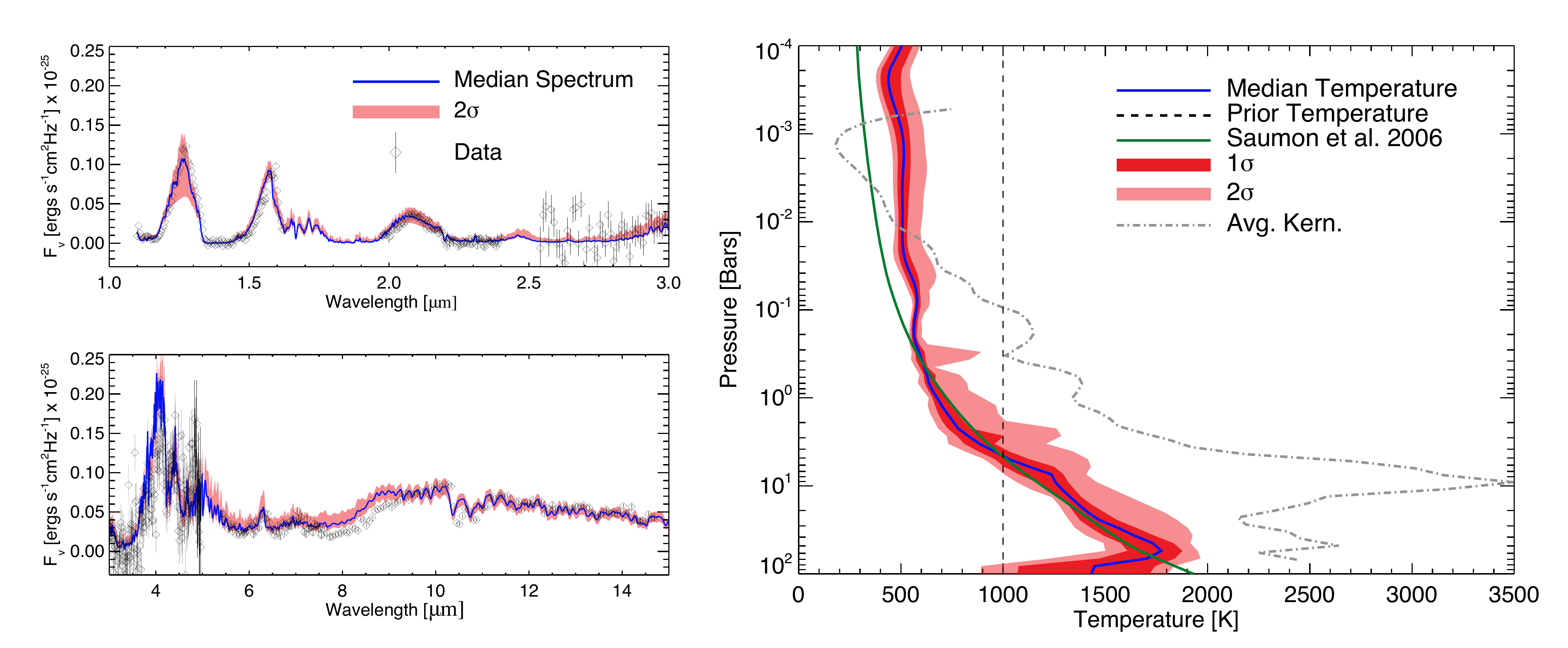}
\end{center}
     \caption{ \label{fig:Figure5} Gl 570D spectrum (left) and retrieved temperature profile (right).  The raw data is shown as the black diamonds with error bars.  The blue spectrum is the median spectrum resulting from the fits to $\sim$10000 resampled data sets.  The light red shaded region shows the 2- $\sigma$ spread in the spectra.     The temperature profile retrieval results are summarized with the median of the ensemble of fits (blue), and the 1- and 2- $\sigma$ spread in the profiles (dark and light red).  The vertical dashed line is the prior mean temperature profile used to initiate each of the fits.  The shaded dot-dashed curve is the normalized averaging kernel for a single fit that shows the pressure levels over which the data is sensitive.  The green curve is the best fit temperature profile from Saumon et al. (2006).   Note the good agreement over the region encompassed by the the averaging kernel.  The model does a poor job between 8 and 9$\mu$m.  This can be remedied by increasing the methane abundance, however the over all fit statistic gets worse due to a stronger absorption in the SpeX and AKARI data.   }
\end{figure*}

\begin{figure*}
\begin{center}
\includegraphics[width=0.75\textwidth, angle=0]{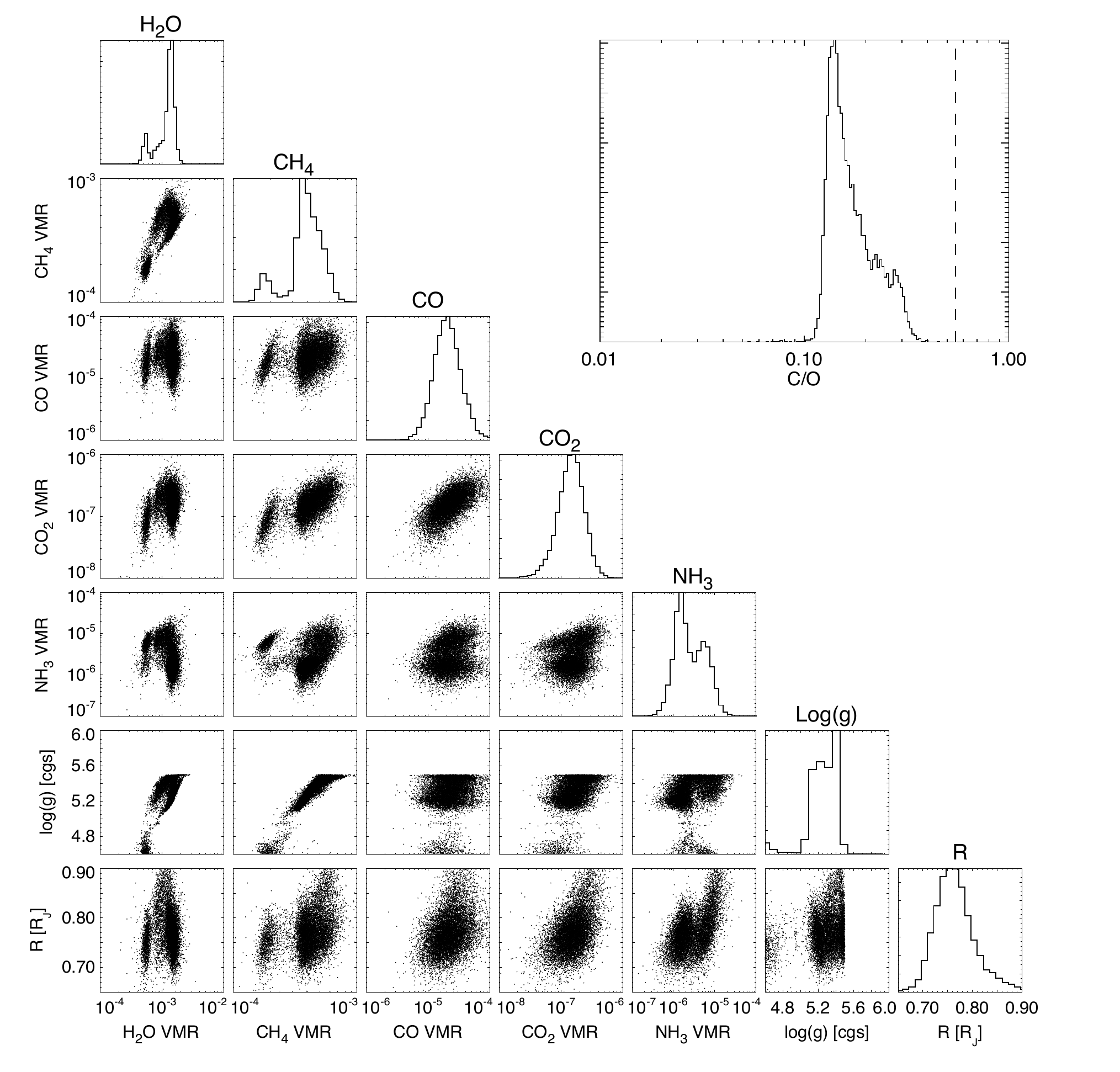}
\end{center}
     \caption{ \label{fig:Figure6} Gas, gravity, and radius retrieval uncertainties resulting from the data resampling retrieval approach.  This ``stair-step" plot shows the correlations amongst each pair of parameters.  The gas abundances are given by the log of their volume mixing ratios.  In each cloud of points, one point represents a single fit of the $\sim$10000.  The histograms on the top of each column represent the marginalized uncertainty distribution for that parameter.  Note that there is a hard upper limit on $\log(g)$ of 5.5.  The inset shows the derived C to O ratio distribution resulting from the retrieval uncertainties.   Solar C/O (0.55) is indicated by the dashed line.     }
\end{figure*}

Table \ref{tab:Table1} and Figures \ref{fig:Figure5} - \ref{fig:Figure6} summarize the bootstrap Monte Carlo results.  Figure \ref{fig:Figure5} summarizes the ensemble of fits and temperature profiles.  Since many thousands of spectra are generated we compute the median spectrum (blue) as well as the one (dark red) and two sigma (light red) spread.  The median reduced $\chi^{2}$ (equation \ref{eq:cost_func}) value is $\sim$16.  The resulting temperature profiles are also presented in a similar manner.  The light gray curve is the averaging kernel profile discussed in \S \ref{sec:Methods} for a single representative fit.   The bulk of the information about the atmosphere originates between a few 10's of bars to a 100 mbars.  This is deeper than in our synthetic example due to the higher gravity which pushes the emission level deeper.   A best fit temperature profile from Saumon et al. (2006) (from hereafter, S06) is shown for comparison.  We find excellent agreement within our uncertainties between a few 10's of bars and $\sim$300 mbar.  Again this is the region probed by the observations.  Our profile becomes more isothermal above the 300 mbar level.  This is perhaps due to the prior influencing the retrieval due to a lack of information at these pressures.  However, different isothermal temperature profile priors still result in convergence of the retrieved profile towards this state.   A recent paper by Sorahana et al. (2014) suggested non-LTE processes may play a role in heating the upper atmospheres of brown dwarfs.  This is worth further exploration.

Figure \ref{fig:Figure6} summarizes the gas, radius, and gravity uncertainty distributions and their correlations.  Each point in the cloud represents a single fit to a noise realization.  The more tilted the cloud of points, the stronger the correlation between the two parameters.  The histograms along the diagonal are the marginalized error distributions.   Order-of-magnitude or better constraints on the gas abundances are obtained.   The numerical summary of these results compared to those of S06 are shown in Table \ref{tab:Table1} .  The quoted uncertainty ranges are the marginalized 68\% confidence interval for each parameter.   

Fortney (2012) suggested that T-dwarfs are excellent laboratories for determining the atmospheric carbon-to-oxygen ratio because the infrared opacity is dominated by carbon and oxygen bearing species such as water, methane, and carbon monoxide.  Furthermore, the photospheres of late-T's, as is Gl 570D, are generally un-obstructed by clouds allowing for more reliable identification and determinations of the carbon and oxygen abundances.  Such determinations can be compared to those of stellar C to O ratios (e.g. Nissen 2013; Teske et al. 2013).   As in the synthetic example, we also show the resulting C to O ratio.  The retrieval results suggest a distinctly sub-solar C to O (0.136 - 0.235) ratio by $\sim$7$\sigma$.  However this is not the native C to O ratio as up to $\sim$20\% of the oxygen is tied up in deep silicate clouds.  Adjusting our values for this depletion (assuming oxygen would be equipartitioned amongst the oxygen bearing species) results in lower C/O values between 0.11 and 0.20.  


In general our retrieval results are in good agreement with the bulk parameters derived in S06.   Our retrieved radii are slightly smaller than in S06 and our gravity is slightly larger.  In both cases our uncertainties are comparatively larger.  The retrieved gravity and radii are used to obtain a spectroscopic mass of 35- 74 $M_{\mathrm{J}}$  .  This range more than encompasses the range in S06 (Table \ref{tab:Table1}).  The retrieved range of effective temperatures is obtained by integrating over each of the $\sim$10000 Monte Carlo spectra.  These effective temperatures are consistent with S06 but with a larger uncertainty.   

\begin{figure}
\begin{center}
\includegraphics[width=0.45\textwidth, angle=0]{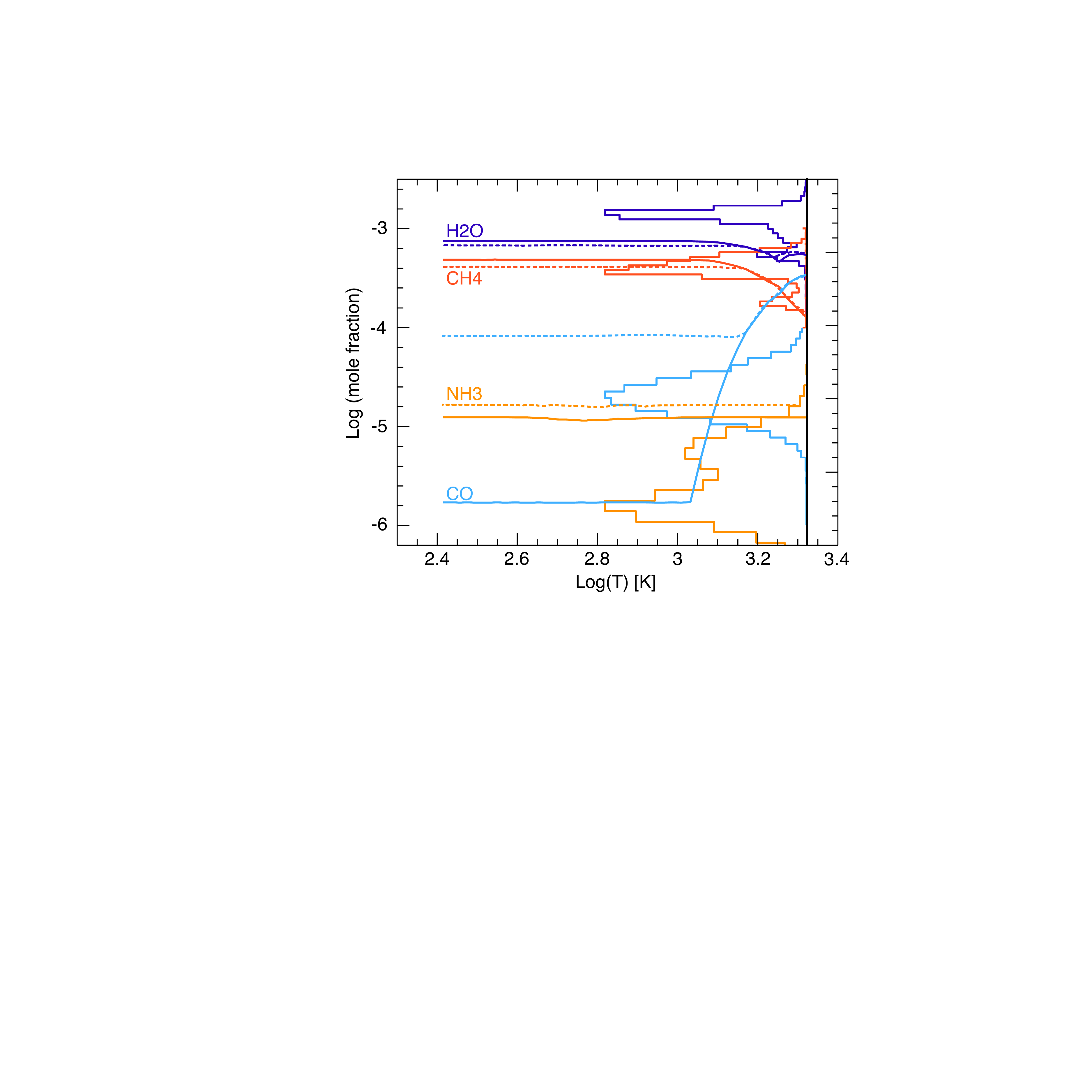}
\end{center}
     \caption{ \label{fig:Figure7} Our abundance retrieval results (histograms) compared with the nominal model of Saumon et al. (2006, solid lines) and Geballe et al. (2009, dashed lines) (adopted from Figure 3 of Saumon et al. (2006) and Figure 5 of Geballe et al. (2009)).  The normalized histograms on the right show the spread in retrieved abundances from the data-resampling retrieval approach (taken from Figure \ref{fig:Figure6}).  Note that they have arbitrary units.}.      
\end{figure}

Figure \ref{fig:Figure7} shows how our abundance results compare with S06 and Geballe et al. (2009).  We show the nominal abundance profiles from Figure 3 of S06 and Figure 5 of Geballe (2009) overlaid with the gas histograms from our retrievals from our Figure \ref{fig:Figure6}.  Our retrieved water abundance distribution is systematically higher, by a factor of $\sim$1.2, than then the S06 nominal profile.  The retrieved methane abundance is in near perfect agreement with S06.  The retrieved CO abundance is larger than their nominal quench level values which is more consistent with deeper thermochemical equilibrium values near $\sim$1260 K.   Geballe et al. (2009) find nearly two orders of magnitude more CO than in S06.  In their investigation M-band spectroscopic data was used to provide better constraints on the CO abundance than what could be done in S06.  In S06 there was no data covering the 4.5 $\mu$m CO fundamental making hard constraints on its abundance difficult.   Our 68\% confidence interval falls between these two cases suggesting an eddy diffusivity, at least as determined by CO alone, between 10$^{2} $cm$^2$ s$^{-1}$ (S06) and 10$^{6} $cm$^2$ s$^{-1}$ (Geballe 2009).  

The largest discrepancy between our retrievals and S06 is the ammonia abundance.  We find nearly an order of magnitude less than S06.  This is difficult to explain with vertical mixing arguments alone due to the nearly uniform abundance of NH$_3$ with increasing depth (see S06).  There have been significant changes in the ammonia line lists (room temperature (Rothman et al. 2005) vs. high temperature (Yurchenko \& Tennyson 2011) ) which may have an impact on the abundance retrieval.  However, S06 present arguments suggesting that changes in the line lists due to the inclusion of hot-bands are unlikely to strongly affect impact the spectra as well as Saumon et al. (2012) demonstrating that the change in cross sections has a minimal impact on the spectra.   We do not yet have a good physical explanation as to why the retrieval is systematically lower than S06.  Perhaps the largest difference in our approach verses those of S06 and Geballe et al. (2009) is their use of {\it both} spectral and evolution model information to find the most self-consistent fit.  Those investigations identified model spectra by using the bolometric luminosity combined with the evolution models to constrain the effective temperature, gravity, and radius.  Our approach uses no {\it a priori} evolutionary information and is just purely driven by the spectral information alone.

\section{Discussion \& Conclusions}\label{sec:Conclusions}
We have for the first time applied well proven, data driven, temperature and abundance retrieval techniques to a brown dwarf spectrum free from many of the physical assumptions present in grid models.  These approaches allow an unbiased determination of the temperatures and abundances in brown dwarf atmospheres.  We first demonstrated that the optimal estimation retrieval approach is a powerful atmospheric inference tool in the presence of gaussian uncertainties.  With a model that is will matched to the data and minimal systematic uncertainties for typical observational conditions, abundances in brown dwarf atmospheres can be determined to within a few tens of percent, compared with orders-of-magnitude for exoplanet data.  Furthermore, the full temperature profile can be constrained at most pressure levels to better than 50 K.  

We then applied our approach to the well studied brown dwarf, Gl 570D.  For the first time,  by combining SpeX, AKARI, and Spitzer IRS data we were able to constrain the abundances of water, methane, carbon monoxide, carbon dioxide, and ammonia as well as temperature structure, gravity and radius (and hence mass).  Real data is plagued with systematic uncertainties due to photometric calibration errors and missing forward model physics.  We were able to at least accommodate for the photometric uncertainties with the bootstrap Monte Carlo approach.   In lieu of these systematic uncertainties we were able to constrain the molecular mixing ratios to better than order-of-magnitude precisions, the temperature profile to within $\sim$100 K at most pressure levels, the effective temperature to within $\sim$50 K, and the mass to within a factor of two.  We found that our results are fairly consistent with those of Saumon et al. (2006) with the largest difference being a lower retrieved ammonia abundance.

As with any data-model comparison method, our approach does have some shortcomings and limitations.  First and foremost, we do not test to see if the derived thermal profile with the derived abundances are in radiative equilibrium in the upper part of the atmosphere. It may be that our solution would be expected to relax rapidly to a different thermal state.   We simply allow the data to guide the solutions rather than our preconceived notions on how the atmosphere {\em should} behave.  Therefore within this weakness lies our potential strength in the sense that we are not bound by these assumptions; we can test whether or not such assumptions are correct. For instance, by not assuming radiative equilibrium we allow for other possible heating mechanisms, such as gravity wave breaking (Young et al. 1997) or non-LTE heating processes (Sorahana et al. 2014), not commonly accounted for within the standard grid modeling framework.  Another weakness, with the current forward model at least, is the assumption of abundances that are uniform with height.  This assumption and whether or not the data justifies including more complicated abundance profiles can be tested in the future.  Additionally, we have not included clouds, since for this object we are comparing results from previous cloud-free grid models (e.g., Saumon et al. 2006; Geballe et al. 2009).  We may also be limited by the accuracy of the cross section databases.  These short comings fall under the more general category of``missing model physics" which when properly accounted for (a task for future investigations) may result in an overall increase the uncertainties on the desired model parameters.  Another potential issue is the applicability of our approach to lower quality data.  In these scenarios the gaussian posterior assumption used in the optimal estimation approach may not be valid. Line et al. (2013) explore in full detail when this assumption is valid, and when more sophisticated techniques (e.g., bootstrap monte carlo and Markov chain Monte Carlo) are necessary.

Finally, there is a continuum of approaches with which one may use in the atmospheric retrieval problem.  On one end, a simple forward model with few limiting assumptions can be used, as was done in this investigation.  On the other end is the self-consistent grid modeling approach where by many assumptions are made to reduce the problem to just a few simple parameters.  It is worth while in future investigations to explore this continuum and to understand how differing physical assumptions can change the results.  Our philosophy is to start with the minimal number of assumptions and build in layers of sophistication as we go.  

There are many questions that we can begin to ask and potentially answer with our retrieval approach. What is the atmospheric temperature structure of Brown Dwarfs? Are they in radiative 
equilibrium as current theory suggests? Can deviations from radiative equilibrium be a potential diagnostic for dynamical processes?  What are the compositions of the brown dwarfs?  Do brown dwarfs have non-solar elemental ratios? In other words, is the metallicity enhancement different for each element? Does composition vary with altitude? How do they deviate from thermochemical equilibrium? What is the vigor of vertical mixing?  Are the observations even sensitive to vertical variations in the abundance profiles?  How will clouds impact the retrievals?  Can we determine the cloud opacities and cloud levels?  Is there missing physics in the self consistent models and can we aid in identification in that missing physics?  Are the observations sensitive to thermal variations in the temperature profile as a function of time?  Are the elemental compositional differences between different objects?  Does this inform us on their formation environments?  The unprecedented quality of brown dwarf data will allow us to begin to address many of these tantalizing questions.

\section{Acknowledgements}		
We thank Didier Saumon, Caroline Morley, and the anonymous for reading the manuscript and providing instructive comments.    JJF acknowledges the suppers of NSF grant AST-1312545.  MSM acknowledges support of the NASA Astrophysics Theory Program.


\begin{thebibliography}

\bibitem[Allard et al.(1996)]{1996ApJ...465L.123A} Allard, F., Hauschildt, P.~H., Baraffe, I., \& Chabrier, G.\ 1996, \apjl, 465, L123 
\bibitem[Allard et al.(1997)]{1997ARA&A..35..137A} Allard, F., Hauschildt, P.~H., Alexander, D.~R., \& Starrfield, S.\ 1997, \araa, 35, 137 
\bibitem[Allard et al.(2001)]{2001ApJ...556..357A} Allard, F., Hauschildt, P.~H., Alexander, D.~R., Tamanai, A., \& Schweitzer, A.\ 2001, \apj, 556, 357 
\bibitem[Barstow et al.(2013)]{2013MNRAS.434.2616B} Barstow, J.~K., Aigrain, S., Irwin, P.~G.~J., Fletcher, L.~N., \& Lee, J.-M.\ 2013, \mnras, 434, 2616 
\bibitem[Burgasser et al.(2000)]{2000ApJ...531L..57B} Burgasser, A.~J., Kirkpatrick, J.~D., Cutri, R.~M., et al.\ 2000, \apjl, 531, L57 
\bibitem[Burgasser et al.(2003)]{2003ApJ...594..510B} Burgasser, A.~J., Kirkpatrick, J.~D., Liebert, J., \& Burrows, A.\ 2003, \apj, 594, 510
\bibitem[Burgasser et al.(2006)]{2006ApJ...637.1067B} Burgasser, A.~J., Geballe, T.~R., Leggett, S.~K., Kirkpatrick, J.~D., \& Golimowski, D.~A.\ 2006, \apj, 637, 1067 
\bibitem[Burgasser et al.(2007)]{2007ApJ...657..494B} Burgasser, A.~J., Cruz, K.~L., \& Kirkpatrick, J.~D.\ 2007, \apj, 657, 494
\bibitem[Burrows et al.(1993)]{1993ApJ...406..158B} Burrows, A., Hubbard, W.~B., Saumon, D., \& Lunine, J.~I.\ 1993, \apj, 406, 158 
\bibitem[Burrows et al.(2006)]{2006ApJ...640.1063B} Burrows, A., Sudarsky, D., \& Hubeny, I.\ 2006, \apj, 640, 1063 
\bibitem[Crisp et al. (2004)]{2004Crisp} Crisp, D., Atlas, R. M., Breon, F. M., Brown, L. R., Burrows, J. P., Ciais, P., Schroll, S. \ 2004,  Adv. in Space Res., 34, 700-709.
\bibitem[Conrath et al.(1998)]{1998Icar..135..501C} Conrath, B.~J., Gierasch, P.~J., \& Ustinov, E.~A.\ 1998, Icarus, 135, 501
\bibitem[Cushing et al.(2008)]{2008ApJ...678.1372C} Cushing, M.~C., Marley, M.~S., Saumon, D., et al.\ 2008, \apj, 678, 1372 
\bibitem[Fletcher et al.(2007)]{2007Icar..189..457F} Fletcher, L.~N., Irwin, P.~G.~J., Teanby, N.~A., et al.\ 2007, Icarus, 189, 457 
\bibitem[Fortney(2012)]{2012ApJ...747L..27F} Fortney, J.~J.\ 2012, \apjl, 747, L27
\bibitem[Geballe et al.(2001)]{2001ApJ...556..373G} Geballe, T.~R., Saumon, D., Leggett, S.~K., et al.\ 2001, \apj, 556, 373
\bibitem[Geballe et al.(2009)]{2009ApJ...695..844G} Geballe, T.~R., Saumon, D., Golimowski, D.~A., et al.\ 2009, \apj, 695, 844 
\bibitem[Greathouse et al.(2011)]{2011Icar..214..606G} Greathouse, T.~K., Richter, M., Lacy, J., et al.\ 2011, Icarus, 214, 606
\bibitem[Helling et al.(2014)]{2014arXiv1403.4420H} Helling, C., Woitke, P., Rimmer, P.~B., et al.\ 2014, arXiv:1403.4420 
\bibitem[Houck et al.(2004)]{2004SPIE.5487...62H} Houck, J.~R., Roellig, T.~L., Van Cleve, J., et al.\ 2004, \procspie, 5487, 62
\bibitem[Hubeny \& Burrows(2007)]{2007ApJ...669.1248H} Hubeny, I., \& Burrows, A.\ 2007, \apj, 669, 1248 
\bibitem[Irwin et al.(2008)]{2008JQSRT.109.1136I} Irwin, P.~G.~J., Teanby, N.~A., de Kok, R., et al.\ 2008, \jqsrt, 109, 1136
\bibitem[Lee et al.(2012)]{2012MNRAS.420..170L} Lee, J.-M., Fletcher, L.~N., \& Irwin, P.~G.~J.\ 2012, \mnras, 420, 170 
\bibitem[Lee et al.(2013)]{2013ApJ...778...97L} Lee, J.-M., Heng, K., \& Irwin, P.~G.~J.\ 2013, \apj, 778, 97
\bibitem[Leggett et al.(2002)]{2002ApJ...564..452L} Leggett, S.~K., Golimowski, D.~A., Fan, X., et al.\ 2002, \apj, 564, 452
\bibitem[Line et al.(2012)]{2012ApJ...749...93L} Line, M.~R., Zhang, X., Vasisht, G., et al.\ 2012, \apj, 749, 93 
\bibitem[Line et al.(2013)]{2013ApJ...775..137L} Line, M.~R., Wolf, A.~S., Zhang, X., et al.\ 2013, \apj, 775, 137 
\bibitem[Line et al.(2014)]{2014ApJ...783...70L} Line, M.~R., Knutson, H., Wolf, A.~S., \& Yung, Y.~L.\ 2014, \apj, 783, 70 
\bibitem[Madhusudhan \& Seager(2009)]{2009ApJ...707...24M} Madhusudhan, N., \& Seager, S.\ 2009, \apj, 707, 24 
\bibitem[Marley et al.(1996)]{1996Sci...272.1919M} Marley, M.~S., Saumon, D., Guillot, T., et al.\ 1996, Science, 272, 1919
\bibitem[Morley et al.(2012)]{2012ApJ...756..172M} Morley, C.~V., Fortney, J.~J., Marley, M.~S., et al.\ 2012, \apj, 756, 172
\bibitem[Moses et al.(2011)]{2011ApJ...737...15M} Moses, J.~I., Visscher, C., Fortney, J.~J., et al.\ 2011, \apj, 737, 15 
\bibitem[{\"O}berg et al.(2011)]{2011ApJ...743L..16O} {\"O}berg, K.~I., Murray-Clay, R., \& Bergin, E.~A.\ 2011, \apjl, 743, L16 
\bibitem[Patience et al.(2012)]{2012A&A...540A..85P} Patience, J., King, R.~R., De Rosa, R.~J., et al.\ 2012, \aap, 540, A85 
\bibitem[Perryman et al.(1997)]{1997A&A...323L..49P} Perryman, M.~A.~C., Lindegren, L., Kovalevsky, J., et al.\ 1997, \aap, 323, L49 
\bibitem[Nissen(2013)]{2013A&A...552A..73N} Nissen, P.~E.\ 2013, \aap, 552, A73 
\bibitem[Nixon et al.(2007)]{2007Icar..188...47N} Nixon, C.~A., Achterberg, R.~K., Conrath, B.~J., et al.\ 2007, Icarus, 188, 47 
\bibitem[Patten et al.(2006)]{2006ApJ...651..502P} Patten, B.~M., Stauffer, J.~R., Burrows, A., et al.\ 2006, \apj, 651, 502 
\bibitem[Rayner et al.(2003)]{2003PASP..115..362R} Rayner, J.~T., Toomey, D.~W., Onaka, P.~M., et al.\ 2003, \pasp, 115, 362
\bibitem[Rice et al.(2010)]{2010ApJS..186...63R} Rice, E.~L., Barman, T., Mclean, I.~S., Prato, L., \& Kirkpatrick, J.~D.\ 2010, \apjs, 186, 63
\bibitem[Rodgers(1976)]{1976RvGSP..14..609R} Rodgers, C.~D.\ 1976, Reviews of Geophysics and Space Physics, 14, 609 
\bibitem[Rodgers(2000)]{2000SAOPP...2.....R} Rodgers, C.~D.\ 2000, Inverse 
Methods for Atmospheric Sounding - Theory and Practice.~Series: Series on 
Atmospheric Oceanic and Planetary Physics, ISBN: <ISBN>9789812813718</ISBN>.~World 
Scientific Publishing Co.~Pte.~Ltd., Edited by Clive D.~Rodgers, vol.~2, 2
\bibitem[Saumon et al.(2000)]{2000ApJ...541..374S} Saumon, D., Geballe, T.~R., Leggett, S.~K., et al.\ 2000, \apj, 541, 374 
\bibitem[Saumon et al.(2006)]{2006ApJ...647..552S} Saumon, D., Marley, M.~S., Cushing, M.~C., et al.\ 2006, \apj, 647, 552 
\bibitem[Saumon \& Marley(2008)]{2008ApJ...689.1327S} Saumon, D., \& Marley, M.~S.\ 2008, \apj, 689, 1327
\bibitem[Saumon et al.(2012)]{2012ApJ...750...74S} Saumon, D., Marley, M.~S., Abel, M., Frommhold, L., \& Freedman, R.~S.\ 2012, \apj, 750, 74 
\bibitem[Sorahana \& Yamamura(2012)]{2012ApJ...760..151S} Sorahana, S., \& Yamamura, I.\ 2012, \apj, 760, 151
\bibitem[Sorahana et al.(2014)]{2014arXiv1401.5801S} Sorahana, S., Suzuki, T.~K., \& Yamamura, I.\ 2014, arXiv:1401.5801 
\bibitem[Stephens et al.(2009)]{2009ApJ...702..154S} Stephens, D.~C., Leggett, S.~K., Cushing, M.~C., et al.\ 2009, \apj, 702, 154 
\bibitem[Teske et al.(2013)]{2013ApJ...778..132T} Teske, J.~K., Cunha, K., Schuler, S.~C., Griffith, C.~A., \& Smith, V.~V.\ 2013, \apj, 778, 132
\bibitem[Tsuji et al.(2011)]{2011ApJ...734...73T} Tsuji, T., Yamamura, I., \& Sorahana, S.\ 2011, \apj, 734, 73 
\bibitem[Twomey et al.(1977)]{1977JAtS...34.1085T} Twomey, S., Herman, B., \& Rabinoff, R.\ 1977, JAS, 34, 1085 
\bibitem[Visscher \& Moses(2011)]{2011ApJ...738...72V} Visscher, C., \& Moses, J.~I.\ 2011, \apj, 738, 72 
\bibitem[Yamamura et al.(2010)]{2010ApJ...722..682Y} Yamamura, I., Tsuji, T., \& Tanab{\'e}, T.\ 2010, \apj, 722, 682
\bibitem[Young et al.(1997)]{1997Sci...276..108Y} Young, L.~A., Yelle, R.~V., Young, R., Seiff, A., \& Kirk, D.~B.\ 1997, Science, 276, 108 


\end{thebibliography}
\end{document}